\documentclass[a4paper,12pt]{article}
\usepackage{amsmath,amsfonts,amssymb,cite}
\newcommand{\be}{\begin{equation}}
\newcommand{\ee}{\end{equation}}

\begin{document}

\begin{center}
{\Large\bf Regge spectrum from holographic models inspired by OPE}
\end{center}

\begin{center}
{\large S. S. Afonin\footnote{On leave of absence from V. A. Fock
Department of Theoretical Physics, St. Petersburg State
University, 1 ul. Ulyanovskaya, 198504, Russia. E-mail:
afonin24@mail.ru.}}
\end{center}

\begin{center}
{\it Institute for Theoretical Physics II, Ruhr University Bochum,
150 Universit\"{a}tsstrasse, 44780 Bochum, Germany}
\end{center}

\begin{abstract}
The problem of obtaining the Regge-like behaviour for meson mass
spectrum in the hard-wall AdS/QCD models is addressed. We show
that the problem can be solved by a simple incorporation of the
effects of local vacuum condensates into such models. The slope
of trajectories turns out to be determined by the local condensate
of dimension~2 that is absent in the standard Operator Product
Expansion. This pitfall, however, can be escaped by means of
physically plausible modification of boundary conditions for the
holographic fields corresponding to the usual gluon condensate,
the latter then determines the slope.
\end{abstract}

\section{Introduction}

Recently the bottom-up holographic models have experienced
noticeable success in description of non-perturbative QCD (see,
e.g.,~\cite{son1,son2,br,pom}). In particular, many important
aspects of chiral dynamics can be reproduced within the simplest
hard-wall model~\cite{son1}. However, the spectrum of highly
excited mesons in the hard-wall models~\cite{son1,br,pom} behaves
like $m_n\sim n$ that does not agree with the Regge-type spectrum,
$m_n^2\sim n$, expected in QCD. To solve this problem one resorted
to an additional input from the 5d side --- the quadratic dilaton
background (the so-called soft-wall model~\cite{son2}). Such an
{\it ad hoc} solution, however, lacks for a simple description of
the chiral symmetry breaking (see, e.g., the recent discussions
in~\cite{gh}). In addition, the physical meaning of the dilaton
background is completely obscure. The slope of trajectories
determines the physical masses of hadrons, hence, the slope itself
is determined by the confinement. The authors of Ref.~\cite{son2}
suggested that the quadratic dilaton could reflect the closed
string tachyon condensation since the latter process is often
believed to be dual to the confinement in the gauge theories. It
was further speculatively assumed in~\cite{csaki} that a closed
string tachyon could be associated holographically to a
dimension~2 condensate whose relation to various aspects of
non-perturbative physics had been widely discussed in the
literature (see, e.g.,~\cite{zak}). This conjecture opens
the doors to a more general idea that, instead of "playing" with
metric, some important aspects of the confinement can be
incorporated into the AdS/QCD models through the account for the
local condensates. An essential feature in this program is that
the local condensates must be taken into consideration
dynamically, this is opposite to the geometric approach
of~\cite{hirn} where the condensates were included through a
modification of metric.

In the present Letter, we propose a very simple and straightforward
way of incorporation of local condensates into the AdS/QCD models.
We make the case of the vector mesons within the hard-wall
holographic model on the simplest Randall-Sundrum background (a
somewhat similar but much more complicated analysis was performed
in~\cite{csaki} for the glueball spectrum). It is shown that with
this theoretical setup the spectrum becomes Regge-like at not
very high energies, with the slope of trajectories being
determined by the dim2 condensate. The latter circumstance is
troublesome because such a gauge-invariant local condensate is
absent in the standard Operator Product Expansion
(OPE)~\cite{svz}. If we regard this condensate as non-local then
we will be in conflict with the AdS/CFT correspondence principle:
The holographic fields must be associated to local gauge-invariant
operators in the field theory. We have found a simple and
intuitively plausible solution of this problem --- the slope can
be related to the usual dim4 gluon condensate by a certain
modification of boundary conditions for the corresponding dual
holographic field, therefore there is no need for introducing the
dim2 condensate.

The paper is organized as follows. The general scheme is presented
in Section~2. In Section~3, we consider a concrete model
demonstrating the underlying idea. Section~4 is devoted to
discussions where it is shown how the problem with dim2 condensate
can be avoided. Finally we provide concluding remarks in
Section~5.

\section{General scheme}

We will demonstrate the idea for the case of isoscalar vector
mesons ($\omega$-mesons), this case can be straightforwardly
generalized to other kinds of mesons. To select the operators that
are important in the formation of masses of resonances we will be
guided by the OPE for the two-point
correlators of quark currents at large Euclidean momentum $Q$. It
has the following structure~\cite{svz} (we omit the Lorentz indices and
irrelevant factors)
\be
\label{1}
\Pi(Q^2)=C_0\log\frac{\Lambda^2}{Q^2}+\sum_{k=1}^{\infty}C_{k}\frac{\langle\mathcal{O}_{2k}\rangle}{Q^{2k}},
\ee
here $\mathcal{O}_{2k}$ are local gauge-invariant operators of canonical
dimension $2k$ (several operators can correspond to the same $k$) and $C_k$
are constants that can be calculated by means of the perturbation theory.
The vacuum expectation values (v.e.v.) of the operators in the r.h.s. of Eq.~\eqref{1}
determine the masses of mesons,
this is especially evident in the large-$N_c$ limit of QCD, where the l.h.s.
of Eq.~\eqref{1} represents a sum over meson poles. As is usual in calculation
of masses within the QCD sum rules~\cite{svz} we should retain only the first several
contributions to the r.h.s. of Eq.~\eqref{1}.

Let us discuss the relevant operators. The first term in the
r.h.s. of Eq.~\eqref{1} is perturbative, hence, it is not of
interest for us. The local gauge invariant operator of dimension~2
is absent in the standard OPE~\cite{svz}, nevertheless, as is
widely discussed in the literature (see, e.g.,~\cite{zak}), an
effective formation of dim2 gluon condensate $\langle
A_{\mu}^2\rangle$ may turn out to be of high importance in the
gluodynamics and lead to a rich phenomenology. In addition, such a
quadratic correction is often associated with contributions of
renormalons (see, e.g., discussions in~\cite{zak2}). For this
reason we tentatively include this operator into our analysis. The
v.e.v. $\langle\mathcal{O}_4\rangle$ is contributed by
$m_q\langle\bar{q}q\rangle$ and by the gluon condensate
$\alpha_s\langle G_{\mu\nu}^2\rangle$. The both have no anomalous
dimension (the last in one loop). The relative contribution of
$m_q\langle\bar{q}q\rangle$ is very small, we neglect it adopting
the chiral limit $m_q=0$. The physical meaning of higher terms in
the asymptotic expansion~\eqref{1} is not essential for our
purposes.

According to the AdS/CFT correspondence, we must link each
operator $\mathcal{O}(x)$ to a field $\varphi(x,z)$ in the 5d bulk
theory, with the 5d masses of fields $\varphi(x,z)$ being
determined via the relation~\cite{gub,wit}
\be
\label{2}
m_5^2=(\Delta-p)(\Delta+p-4),
\ee
where $\Delta$ is the canonical dimension of the corresponding
operator $\mathcal{O}(x)$ and $p$ in our simple case is just the number
of Lorentz indices.

Let us discuss the field content in our model. First of all, we
have the vector meson that is interpolated by the current
$\bar{q}\gamma_{\mu}q$. The corresponding 5d field
will be denoted $V_M$. We have $p=1$, $\Delta=3$, hence,
$(m_5)_{V}^2=0$. Second, we have an infinite set of operators
$\langle\mathcal{O}_{2k}\rangle$ (for simplicity, we neglect the
fact that, generally speaking, a finite number of different
operators corresponds to each $k$). The corresponding 5d scalar
fields $X_{2k}$ have $p=0$, $\Delta=2k$, hence,
\be
\label{2b}
(m_5)_{2k}^2=4k(k-2).
\ee

The action of the theory in the bulk describing the vector mesons is
\be
\label{3}
S=\int d^4\!x\,dz\sqrt{g}\,\text{Tr}\left\{\sum_k\left(|DX_{2k}|^2-(m_5)_{2k}^2|X_{2k}|^2\right)
-\frac14F_{MN}^2\right\},
\ee
where
\be
\label{4}
D_MX_{2k}=\partial_MX_{2k}-ig_5V_MX_{2k},
\ee
\be
\label{5}
F_{MN}=\partial_MV_N-\partial_NV_M.
\ee
As usual, the holographic coordinate $z$ corresponds to the inverse
energy scale, $z\sim1/Q$. The v.e.v. of the fields $X$ are determined
by the classical solutions satisfying the UV boundary conditions.
Since in the limit of very high energies the vacuum in QCD is perturbative,
i.e. there are no condensates, it is natural to impose the UV boundary condition
\be
\label{6}
X_{2k}(x,z=0)=0.
\ee
At this stage we do not impose any IR boundary (to be discussed below)
that determines a scale
until which the running of the QCD gauge coupling is neglected. We believe
that in dealing with renorminvariant (or almost renorminvariant) quantities,
as we do, the problem of running coupling is irrelevant.

To obtain a concrete model one has to choose a metric, the
classical solutions for $\langle X_{2k}\rangle$ form then a
"potential" for the vector field $V_M$, fixing for the latter a
gauge and boundary conditions one calculates the mass spectrum.

\section{A model}

We will consider the simplest metric exploited in the hard-wall
model, the anti-de Sitter one~\cite{son1},
\be
\label{7}
ds^2=\frac{1}{z^2}(dx_{\mu}dx^{\mu}-dz^2),
\ee
where for simplicity it is taken $R=1$ for the radius of AdS space.
Fixing the gauge $V_z=0$, the normalized solutions $v_n$,
$V_M^n(x,z)=V_{\mu}^n(x)v_n(z)$, of classical
equation for the transverse components $V_{\mu}^T$ exist only for
discrete values of 4d momentum $q^2=m_n^2$,
\be
\label{8}
\partial_z\left(\frac{1}{z}\partial_zv_n\right)+\frac{m_n^2v_n}{z}=
\frac{2g_5^2}{z^3}v_n\sum_k\langle X_{2k}\rangle^2,
\ee
where $\langle X_{2k}\rangle$ are solutions of
\be
\label{9}
\frac{1}{z^3}\partial_{\mu}\partial^{\mu}X_{2k}-\partial_z\left(\frac{1}{z^3}\partial_zX_{2k}\right)=
-\frac{1}{z^5}(m_5)_{2k}^2X_{2k},\qquad k=1,2,\dots
\ee
We look for solutions for $\langle X_{2k}\rangle$ which are functions
of $z$ only. Making change of variables $v_n=\sqrt{z}\psi_n$ the system
of equations~\eqref{8},~\eqref{9} takes the form
\be
\label{10}
-\psi_n''+\left(\frac{3}{4z^2}+\frac{2g_5^2}{z^2}\sum_k\langle X_{2k}\rangle^2\right)\psi_n=
m_n^2\psi_n,
\ee
\be
\label{11}
z^2X_{2k}''-3zX_{2k}'-(m_5)_{2k}^2X_{2k}=0,\qquad k=1,2,\dots
\ee

Equation~\eqref{10} is of Schr\"{o}dinger type with the "potential" in brackets,
this potential is determined by the solutions of Eqs.~\eqref{11}. Inserting
the values of $(m_5)_{2k}^2$ from relation~\eqref{2b} into Eqs.~\eqref{11}
we obtain the following solutions satisfying the boundary condition~\eqref{6},
\be
\label{12}
\langle X_2\rangle=c_2^{(1)}z^2+c_2^{(2)}z^2\log z,
\ee
\be
\label{13}
\langle X_{2k}\rangle=c_{2k}^{(1)}z^{2k},\qquad k=2,3,\dots,
\ee
where $c_{2k}^{(i)}$ are some dimensional constants.

In order to have the Regge-like spectrum, $m_n^2\sim n$, the potential in
Eq.~\eqref{10} has to be of oscillator type, i.e. to behave as $z^2$ at large $z$.
This is achieved if we set $c_2^{(2)}=0$ and neglect the contributions of
higher-dimensional operators, $c_{2k}^{(1)}=0$ at $k>1$. The spectrum will be
\be
\label{14}
m_n^2=4\sqrt{2}\,g_5|c_2^{(1)}|n+\text{const}.
\ee
In particular, choosing $c_2^{(1)}=(\sqrt{2}\,g_5)^{-1}$ in appropriate units of
energy square we arrive at the spectrum obtained
in the simplest soft-wall model~\cite{son2}. We note also that since the slope
$a=4\sqrt{2}\,g_5|c_2^{(1)}|$ in Eq.~\eqref{14} is proportional to the string
tension $\sigma$, $a=2\pi\sigma$, the quadratic correction in the OPE~\eqref{1}
turns out to be also proportional to $\sigma$, this is in a qualitative agreement
with the results of holographic calculations performed in~\cite{and}.

Thus, within the given holographic model, the first non-perturbative contribution
to the two-point correlators, the so-called dim2 gluon condensate, is responsible for
the Regge-like behaviour of meson spectrum, the higher non-perturbative contributions
in the OPE~\eqref{1} yield anharmonic corrections to the spectrum.

\section{Discussions}

In a sense, the principle of AdS/CFT correspondence converts the
asymptotic expansion in $Q^{-2}$ in the OPE into asymptotic
expansion in $z^4$ in the holographic potential of equation for
mass spectrum~\eqref{10},
\be
\label{14b}
U(z)=\frac{3}{4z^2}+2g_5^2\sum_{k=1}^{\infty}c_kz^{4k-2}.
\ee
In the QCD sum rules~\cite{svz}, only the lowest contributions in the
OPE are essential for the determination of masses of ground states
of mesons. Accepting the same principle in the model above we
should also neglect the higher power-like contributions to the
spectrum.

Until now we have nothing said about the IR boundary. An effective
IR boundary $z_{\text{IR}}$ should certainly exist in the
presented model, but in contrast to the hard-wall models we do not
impose any special boundary conditions at $z_{\text{IR}}$ (the
boundary conditions are determined by the requirement to have the
Regge-like spectrum), rather $z_{\text{IR}}$ shows the range of
applicability of the model. For instance, by tuning
$z_{\text{IR}}$ one can achieve the dominance of the
oscillator-type contribution $z^2$ in the potential of
Eq.~\eqref{10} in a certain range of large $z$. Following this way
one obtains a Regge-like spectrum for a certain number of excited
mesons which were observed experimentally, the description of
higher states (not observed experimentally) are then beyond the
validity of the model. More exactly, the shape of potential well
is $\mathcal{O}(z^2)$ at $z_{\text{min}}<z<z_{\text{IR}}$, where
$z_{\text{min}}$ is the minimum of potential, and at
$z=z_{\text{IR}}$ one has a "hard" wall. As a consequence, the
spectrum of normalizable modes is of oscillator type, $m_n^2\sim n$,
at small $n$ and, after imposing the appropriate IR boundary
condition,
represents zeros of Bessel function, $m_n\sim n$, at large $n$,
$m_n^2>U(z_{\text{IR}})$, where the model is supposed to be not
applicable.

The theoretical status of quadratic correction in the
OPE~\eqref{1} is uncertain as long as the existence of dim2 gluon
condensate (the effective "tachyonic" gluon mass) and the related
phenomenology are somewhat speculative presently, let alone the
problems with the implementation of AdS/CFT principle mentioned
in Introduction. The question
appears whether it is possible to modify the presented model such
that the slope of meson trajectories were not determined by the
dim2 gluon condensate? Intuitively it would be more natural to
imagine that the slope is related to the dim4 gluon condensate, is
it possible to implement this? The answer is positive. The dim4
gluon condensate is known to be contributed not only by the
non-perturbative effects but also by the perturbation theory after
summation over certain gluon exchanges. For this reason the UV
boundary condition~\eqref{6} may be just incorrect for the scalar
field corresponding to the dim4 operator $\alpha_s G_{\mu\nu}^2$,
it should be weakened to
\be
\label{15}
X_4(x,z=0)=\text{const}.
\ee
In this case, the solution of Eq.~\eqref{11} for $k=2$ is
\be
\label{16}
\langle X_4\rangle=c_4^{(1)}z^4+c_4^{(2)}.
\ee
Substituting this solution in Eq.~\eqref{10} we observe that
$\langle X_4\rangle$ yields contribution both to the UV (stemming from
$c_4^{(2)}$) and to the IR (stemming from $c_4^{(1)}$)
parts of potential, with the contribution of oscillator
type representing an interplay of the both,
$\mathcal{O}(c_4^{(1)}c_4^{(2)}z^2)$, i.e.
\be
\label{18}
m_n^2\sim c_4^{(1)}c_4^{(2)}n+\dots
\ee
This reflects holographically
the fact that the gluon condensate encodes both perturbative and
non-perturbative effects as we know it from the phenomenology.
In addition, we see a direct realization of the AdS/CFT idea: The UV
behaviour of the 5d dual theory determines the low-energy properties
of the 4d theory on the boundary --- the slope of discrete mass
spectrum in the given case. Now we can remove the dim2
condensate basing our analysis on the standard OPE~\cite{svz}.

It should be noted that the metric can be chosen such that the
Regge slope is automatically determined by the dim4 operators in
the OPE while the dim2 ones do not contribute to the slope even if
they existed. We have found that this happens if the analysis
above is formally performed in the flat metric, see Appendix.

It seems that the description of Regge-like spectrum can be made
fully compatible with the simultaneous description of the chiral
symmetry breaking. If we neglect all contributions that lead to
the anharmonic terms in the potential of Eq.~\eqref{10} and take
into account the contribution of the quark bilinear operator
$\bar{q}q$ in the axial-vector channel following the procedure
described in~\cite{son1}, this contribution will dominate in the
IR-region, on the other hand, the UV asymptotics of the solutions
will be unchanged. Since the description of chiral dynamics is
based on these asymptotics, the corresponding results from~\cite{son1}
seem to be compatible with the simultaneous description of the
Regge-like spectrum.

\section{Concluding remarks}

The experience of holographic models of QCD shows that it is
really hard to achieve a satisfactory comprehensive description of
non-perturbative QCD just by playing with the background metric
and boundary conditions for the fields in the bulk, one should add
some ingredients directly from QCD or low-energy effective
theories of QCD. For instance, knowing that the quark condensate
is the order parameter of the chiral symmetry breaking, one should
introduce a 5d scalar field corresponding to the quark bilinear
operator and organize a nontrivial solution that would correspond
to the v.e.v. of the operator under consideration --- the quark
condensate. This was the first indispensable step for the correct
description of chiral dynamics in~\cite{son1}. In essence, we have
proposed a simple and compact demonstration of the fact that in
order to obtain the correct spectrum of meson excitations it is
not necessary to complicate the bulk geometry, it is sufficient to
add more QCD to the simplest model in a similar way. Thus, the
question "Why does the simplest hard-wall model, being successful
in description of the chiral dynamics, fail to reproduce the
Regge-like spectrum?" has a simple and natural answer: Because the
5d field corresponding to the QCD operator that is crucial for
chiral dynamics --- the quark bilinear operator --- was taken into
account while the 5d fields corresponding to the QCD operators
responsible for the masses of hadrons were not taken into
consideration.

The problem that emerges along this line is the appearance of
anharmonic corrections to the spectrum. As a result, the spectrum
looks like (at least if these corrections are small)
\be
\label{17}
m_n^2\sim\sum_{i=1}^{\infty}c_in^i.
\ee
The role of terms with $i>1$ is an open question. On the one hand,
one may try to elaborate some mechanism for their suppression (or, say,
regard them as an artifact of asymptotic nature of OPE), on the other
hand, the experiment~\cite{pdg} does not provide convincing indications
that they must be suppressed. Within the holographic models, such
anharmonic contributions were systematically analyzed
in~\cite{afonin}. It is quite intriguing to observe that the
solution found for the slope~\eqref{18} shares with the analysis
of~\cite{afonin} the following general feature: The slope is equally
determined by the IR and UV sectors of the underlying theory.

\section*{Acknowledgments}

The work is supported by the Alexander von Humboldt Foundation and
by RFBR, grant 09-02-00073-a. I am grateful for the warm
hospitality by Prof. Maxim Polyakov extended to me at the Bochum
University.

\section*{Appendix}

Consider the flat metric
$$
ds^2=dx_{\mu}dx^{\mu}-dz^2.
\eqno (A1)
$$
In this metric, the classical equations~\eqref{10},~\eqref{11}
take the form
$$
-\psi_n''+\left(2g_5^2\sum_k\langle X_{2k}\rangle^2\right)\psi_n=
m_n^2\psi_n,
\eqno (A2)
$$
$$
X_{2k}''-(m_5)_{2k}^2X_{2k}=0,\qquad k=1,2,\dots
\eqno (A3)
$$
Making use of relation~\eqref{2b} and boundary
condition~\eqref{6}, the solutions of Eqs.~(A3) are
$$
\langle X_2\rangle=c_2\sin(2z),
\eqno (A4)
$$
$$
\langle X_4\rangle=c_4z,
\eqno (A5)
$$
$$
\langle X_{2k}\rangle=c_{2k}\sinh(2\sqrt{k(k-2)}\,z),\qquad k=3,4,\dots
\eqno (A6)
$$
The harmonic contribution to the potential of Eq.~(A2)
stems from the v.e.v. $\langle X_4\rangle$, at least for large enough $z$.
It seems that we should not use our logic for the operators with $k>2$ as
the solutions~(A6) do not look physical.


\begin{thebibliography}{99}
\bibitem{son1} J.~Erlich, E.~Katz, D.~T.~Son and M.~A.~Stephanov,
  Phys. Rev. Lett. {\bf 95}, 261602 (2005).
\bibitem{son2} A.~Karch, E.~Katz, D.~T.~Son and M.~A.~Stephanov,
  Phys. Rev. D {\bf 74}, 015005 (2006).
\bibitem{br} G.~F.~de Teramond and S.~J.~Brodsky,
  Phys. Rev. Lett.  {\bf 94}, 201601 (2005).
\bibitem{pom} L.~Da Rold and A.~Pomarol,
  Nucl. Phys. B {\bf 721}, 79 (2005).
\bibitem{gh} T. Gherghetta, J. I. Kapusta, and T. M. Kelley,
arXiv:0902.1998 [hep-ph].
\bibitem{csaki}  C. Cs\'{a}ki and M. Reece, JHEP {\bf 0705}, 062
(2007).
\bibitem{zak} F. V. Gubarev and V. I. Zakharov,
  Phys. Lett. B {\bf 501}, 28 (2001);
  K. G. Chetirkin, S. Narison, and V. I.~Zakharov,
Nucl. Phys. B {\bf 550}, 353 (1999).
\bibitem{hirn} J. Hirn, N. Rius, and V. Sanz,
  Phys. Rev. D {\bf 73}, 085005 (2006).
\bibitem{svz} M. A.~Shifman, A. I.~Vainstein, and V. I.~Zakharov,
Nucl. Phys. B {\bf 147}, 385, 448 (1979).
\bibitem{zak2} A. I. Vainshtein and V. I.~Zakharov,
    Phys. Rev. Lett. {\bf 73}, 1207 (1994).
\bibitem{gub} S. S. Gubser, I. R. Klebanov, and A. M. Polyakov,
  Phys. Lett. B {\bf 428}, 105 (1998).
\bibitem{wit} E. Witten,
  Adv. Theor. Math. Phys. {\bf 2}, 253 (1998).
\bibitem{and} O. Andreev,
  Phys. Rev. D {\bf 73}, 107901 (2006).
\bibitem{pdg} C. Amsler {\it et al.}, Phys. Lett. B {\bf 667}, 1 (2008).
\bibitem{afonin} S.~S.~Afonin,  Phys. Lett. B {\bf 675}, 54 (2009).


\end{thebibliography}
\end{document}